\documentclass[%
 reprint,
 amsmath,amssymb,
 aps,
]{revtex4-1}

\usepackage{comment}
\usepackage{xcolor,graphicx,float}
\usepackage[normal]{subfigure}
\usepackage{dcolumn}
\usepackage{bm}

\usepackage{bbm}
\usepackage{algpseudocode}

\usepackage{algorithmicx,algorithm}

\usepackage{physics}
\usepackage{hyperref}

\usepackage{mathtools}

\usepackage{amsthm}
\usepackage[capitalise]{cleveref}

\newtheorem{thm}{Theorem}
\crefname{thm}{Theorem}{Theorems}

\crefname{defn}{Definition}{Definitions}

\crefname{lem}{Lemma}{Lemmas}

\crefname{conj}{Conjecture}{Conjectures}

\begin{document}


\title{Algorithms for Finding the Maximum Clique Based on Continuous Time Quantum Walks}
\author{Xi Li}
\email{230169107@seu.edu.cn}
\affiliation{%
School of Cyber Science and Engineering, Southeast University, Nanjing 210096, China}
\author{Hanwu Chen}
\email{hw\_chen@seu.edu.cn}
\affiliation{%
	School of Computer Science and Engineering, Southeast University, Nanjing 210096, China}

\author{Mingyou Wu}
\affiliation{%
School of Cyber Science and Engineering, Southeast University, Nanjing 210096, China}

\begin{abstract}
In this work, the application of continuous time quantum walks (CTQW) to the Maximum Clique (MC) Problem is studied. Performing CTQW on graphs will generate distinct periodic probability amplitudes for different vertices. We found that the intensities of the probability amplitude at some frequencies implies the clique structure of some special kinds of graphs. Recursive algorithms with time complexity $O(N^5)$  in classical computers are proposed for finding the maximum clique. We have experimented on random graphs where each edge exists with probabilities of 0.3, 0.5, and 0.7 . Although counter examples were not found for random graphs, whether these algorithms are universal is beyond the scope of this work.
\end{abstract}

\maketitle

\section{\label{sec:level1}Introduction}
The problem in finding the maximum clique (complete subgraph) is an 
Non-deterministic Polynomial Complete Problem (NP-complete problem)\cite{karp1972reducibility}. The optimal complexity of exact algorithms is $O(2^{0.249N})$\cite{robson2001finding}. Designing a polynomial algorithm for NP-complete problems on classical computers is normally difficult. As the performance of quantum algorithms has been proven to be better than the classical algorithm in most situations\cite{nielsen2002quantum,shor1994algorithms,grover1997quantum}, many scientists have turned to quantum algorithms for NP-complete problems\cite{roland2003adiabatic,cerf2000nested,farhi2014quantum,kaminsky2004scalable,childs2000finding}. With the advantage of quantum states, all possible solutions (combinations of vertices of a given graph) are encoded in an initial superposition state, and the optimal solution is searched by a quantum evolution process in the previous quantum algorithms. The quantum algorithm asymptotically requires the square root of the number of operations that the classical algorithm requires\cite{nielsen2002quantum,roland2003adiabatic}. Apparently, the structure of graphs is not adequately considered in these algorithms. The work of Noga Alon, Michael Krivelevich, and  Benny Sudakov shows that the second eigenvector is related to the MC of random graphs\cite{alon1998finding}. Generally, connecting the structure of graphs with the NP-complete problem is unclear. In this work, the structure of the MC in the graph specifically refers  to whether a vertex belongs to the MC. We mainly focus on the structure of center graphs because all kinds of graphs can be transformed to center graphs. A graph is called a center graph if there exists one so-called center vertex adjacent to all other vertices. We will describe how the clique structure impacts the continuous time quantum walks (CTQW) of several special kinds of center graphs, and algorithms for the maximum clique problem will be proposed. In Farhi and Goldstone’s work, CTQW is defined as an evolution of a quantum system which is driven by the Laplacian matrix of a given graph\cite{farhi1998quantum}. With other physical models \cite{christandl2004perfect,gamble2010two}, the Hamiltonian of the CTQW is defined as the adjacency matrix to the corresponding graph in this work. Then the state of CTQW is determined by $\left| {\varphi \left( t \right)} \right\rangle  = {e^{ iAt}}\left| {\varphi \left( 0 \right)} \right\rangle$, where $A$ is the adjacency matrix of the given graph $G$ and $e^{ iAt}$ is an evolution unitary operator. This operation exists in a series form ${e^{itA}} = \sum\limits_{s = 0}^\infty  {\frac{{{{\left( {it} \right)}^s}{A^s}}}{{s!}}}$. As $A^s$ is characterized by the number of walks in the graphs, the CTQW does reflect the clique structures of several kinds of center graphs. The evolution can be estimated when eigenvectors and eigenvalues of a given adjacency matrix are obtained by numerical computation in $O(N^3)$ time on classical computers. The probability amplitude of CTQW is chosen as the critical feature to infer whether a vertex is a member of the maximum clique.

This paper is organized as follows: the second section presents CTQW on center graphs. In the third section, an algorithm, named algorithm A, with $O(N^5)$ time complexity based on CTQW for finding the maximum clique is introduced. In the fourth section, the probable error of algorithm A is presented and an approach for constructing a graph invalid for algorithm A is described. In the fifth section, we give two algorithms, named algorithm B and algorithm C, to fix problems with algorithm A. The last section is the conclusion.

\section{Clique structure and CTQW on the center graph} 
Generally, a graph is denoted as $G(V,E)$, consisting of a vertex set $V$ and an edge set $E$. The set $E$ is a subset of $V\times V$, which implies the connection relationship between any pair of vertices in $V$. Let the number of $V$ equal to $N$, where the adjacency matrix of $G$ is an $N\times N$ real symmetric matrix $A$, where $A_{jl}=1$ if vertices $v_j$ and $v_l$ are connected, otherwise $A_{jl}=0$.

Consider the CTQW on a given graph, one can associate every vertex $v_j$ of the graph with a basis vector $\left|j\right>$ in an N-dimensional vector space. The Hamiltonian of the system is 
\begin{equation}
H = A,
\end{equation}
If $v_j$ is the initial state of the system, and the transition probability amplitude from $v_j$ to $v_l$ is $\alpha_{l,j}(t)$ or short format $\alpha_{l,j}$, then:

\begin{equation}
{\alpha _{l,j}}\left( t \right) = \left\langle l \right|{e^{ i\gamma At}}\left| j \right\rangle,
\end{equation}
The probability $\pi _{l,j}(t)$ or short format $\pi _{l,j}$ can be written as
\begin{equation}
\pi _{l,j}(t) = \left | \left<l |e^{i \gamma A t} | j\right> \right| ^2.
\end{equation}
The eigenvalues of A are denoted as $\lambda_n$($n=1,2,\ldots,N$), the eigenvalues are arranged in descending order, namely ${\lambda _1} \ge {\lambda _2} \ge  \cdots  \ge {\lambda _N}$. The eigenvector belongs to $\lambda_n$, which is denoted as $\left|\lambda_n\right>$, producing:

\begin{equation}\label{Re}
{\alpha _{l,j}}\left( t \right) = { {\sum\limits_n {{e^{ i {\lambda_n}t}}\left\langle {l}
			\mathrel{\left | {\vphantom {l {{\lambda_n}}}}
				\right. \kern-\nulldelimiterspace}
			{{{\lambda_n}}} \right\rangle \left\langle {{{\lambda_n}}}
			\mathrel{\left | {\vphantom {{{\lambda_n}} j}}
				\right. \kern-\nulldelimiterspace}
			{j} \right\rangle } } },
\end{equation}
and
\begin{equation}
{\pi _{l,j}}\left( t \right) = {\left| {\sum\limits_n {{e^{ i {\lambda_n}t}}\left\langle {l}
			\mathrel{\left | {\vphantom {l {{\lambda_n}}}}
				\right. \kern-\nulldelimiterspace}
			{{{\lambda_n}}} \right\rangle \left\langle {{{\lambda_n}}}
			\mathrel{\left | {\vphantom {{{\lambda_n}} j}}
				\right. \kern-\nulldelimiterspace}
			{j} \right\rangle } } \right|^2}.
\end{equation}
The real part of the amplitude $\alpha _{l,j}$ can be represented as:
\begin{equation}
R\left( {{\alpha _{l,j}}} \right) = \sum\limits_n {{p_n}\cos \left( {{\lambda _n}t} \right)}
\end{equation}
where $p_n= \left\langle {l}
\mathrel{\left | {\vphantom {l {{\lambda_n}}}}
	\right. \kern-\nulldelimiterspace}
{{{\lambda_n}}} \right\rangle \left\langle {{{\lambda_n}}}
\mathrel{\left | {\vphantom {{{\lambda_n}} j}}
	\right. \kern-\nulldelimiterspace}
{j} \right\rangle$. This implies that the real part of the amplitude of CTQW is a periodic function with $N$ frequency components, and the frequency values are the eigenvalues of the adjacency matrix and the intensity of the frequency $\lambda_n$ is $p_n$.

The amplitude can also be represented as a form of sums, i.e.,
\begin{equation}
{\alpha _{l,j}}\left( t \right) = \sum\limits_{s = 0}^\infty  {\frac{{{{\left( {it} \right)}^s}{{\left( {{A^s}} \right)}_{l,j}}}}{{s!}}}.
\end{equation}
where${A^s}_{l,j}$ denotes the number of walks of length $s$ \cite{cvetkovic1980spectra}. Therefore, the CTQW can be determined by the number of walks in the graphs.

Consider a graph $G$ and one of its vertices $v_j$ and let $\textbf{N}(j)$ denote the neighbors of $v_j$. The induced subgraph of vertex $v_j \bigcup \textbf{N}(j)$ is denoted as $G_j$. We call $G_j$ the center subgraph of vertex $v_j$ and the vertex $v_j$ the center vertex of $G_j$. Note that the concept of the center graph is not completely the same as the concept of reference \cite{bonchev1980generalization}. Two natural approaches can be used to transform a graph into a center graph or set of center graphs. The first way is to add a new vertex then connects it to every vertex of the original graph. The second approach is to induce a set of center graphs $\{G_1,\dots,G_N\}$ of the original graph $G$. 

\begin{figure}[!htbp]
	\centering\includegraphics[width=0.45\textwidth]{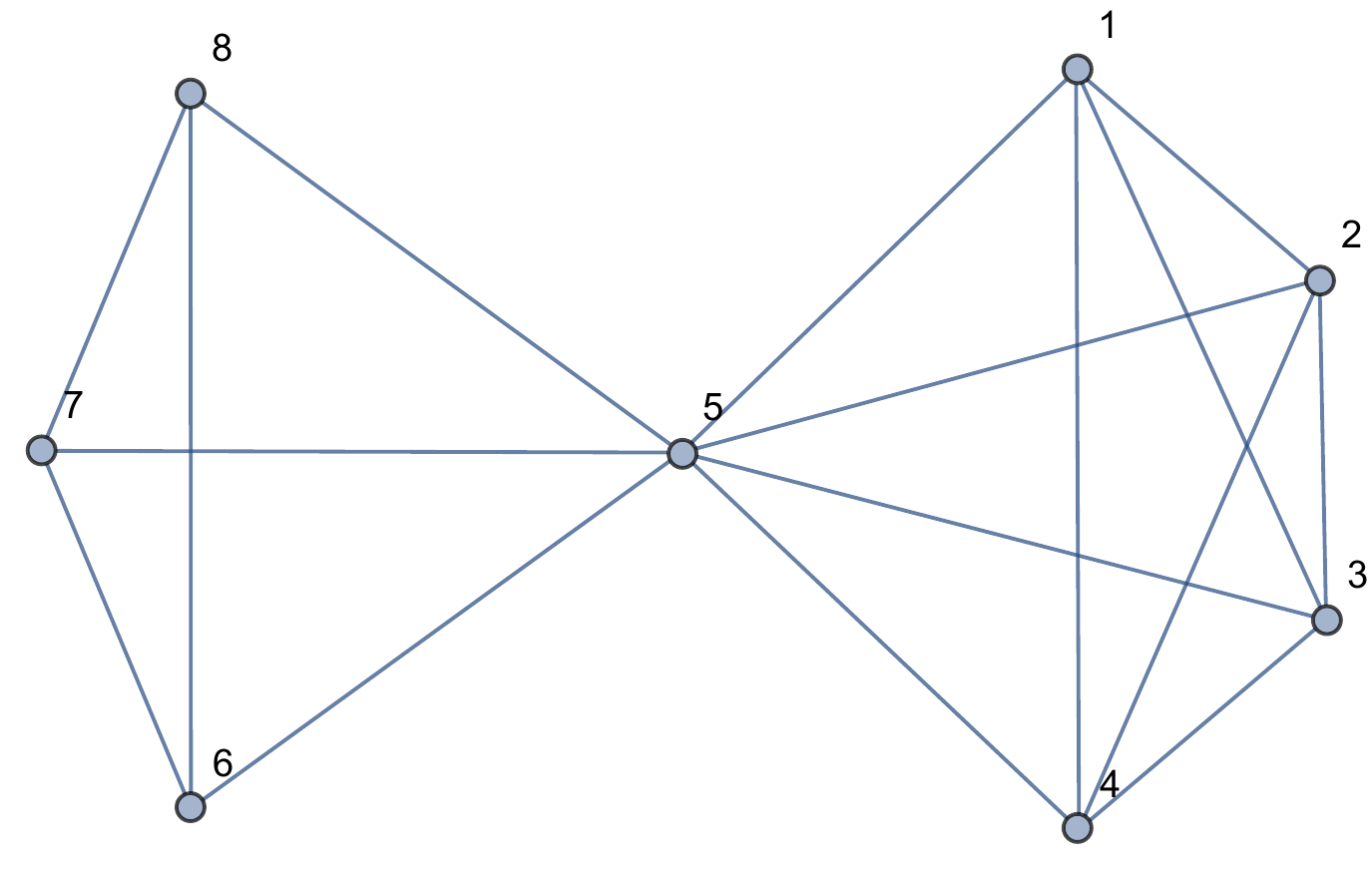}
	\caption{The center graph $G_5$ of vertex $5$. There exists two cliques, one is the subgraph induced by $\{1,2,3,4,5\}$, another is the subgraph induced by $\{5,6,7,8\}$}
	\label{f5}
\end{figure}

A center graph $G_j$ is called the first kind of ideal center graph in this work if there are two cliques in $G_j$ and there is no edge connecting any pair of vertices $\{v_j,v_l\}$ when $v_j$ and $v_l$ are members of distinct cliques. An example of a center graph is shown in Fig.(\ref{f5}).

It is difficult to obtain the analytical solutions of eigenvalues and eigenvectors even for the first kind of ideal center graph. A method of counting the number of walks is used to solve the CTQW. Let $W_s$ also denote the number of closed walks of center vertex, $F_s$ denote the number of walks starting from the center vertex $j$ ending with one of the vertices in the MC, and $H_s$ denote the number of walks starting from the center vertex $j$ ending with one of vertices not in the $M$C. It provides that: 
\begin{equation}\label{eq13}
\left\{ {\begin{array}{*{20}{c}}
	{{W_{s + 1}} = \left( {{m_1} - 1} \right){F_s} + \left( {{m_2} - 1} \right){H_s}}\\
	{{F_{s + 1}} = {W_s} + \left( {{m_1} - 2} \right){F_s}}\\
	{{H_{s + 1}} = {W_s} + \left( {{m_2} - 2} \right){H_s}}
	\end{array}} \right.
\end{equation}
where $m_1$ is the clique number and $m_2$ is the size of the remaining clique. Finding the solutions of Eq.(\ref{eq13}) is equivalent to eigen decomposition of adjacency matrix $A$. However, the exact numerical solutions are complex and do not help to infer whether a vertex belongs to the maximum clique. We only need the relationships between the probability amplitudes of different vertices. In utilizing literature results\cite{cvetkovic1980spectra,van2010graph}, we have 
\begin{equation}\label{eq14}
{W_s} = \sum\limits_{n = 1}^N {{a_n}\lambda _n^s}.
\end{equation}

where $\sum\limits_{n = 1}^N {{a_n}} = 1$. Taking Eq.(\ref{eq14}) into the second and third terms of Eq.(\ref{eq13}), we have 
\begin{equation}\label{eq15}
{F_{s + 1}} = \sum\limits_{n = 1}^N {{a_n}\lambda _n^s}  + \left( {{m_1} - 2} \right){F_s},
\end{equation}
and 
\begin{equation}\label{eq16}
{H_{s + 1}} = \sum\limits_{n = 1}^N {{a_n}\lambda _n^s}  + \left( {{m_1} - 2} \right){H_s}.
\end{equation}

Solving Eq.(\ref{eq15}) and Eq.(\ref{eq16}), we have 
\begin{equation}\label{eq17}
{F_s} = \sum\limits_{n = 1}^N {{a_n}} \frac{{{{\left( {{m_1} - 2} \right)}^s} - \lambda _n^s}}{{{m_1} - 2 - {\lambda _n}}}
\end{equation}
and 
\begin{equation}\label{eq18}
{H_s} = \sum\limits_{n = 1}^N {{a_n}} \frac{{{{\left( {{m_2} - 2} \right)}^s} - \lambda _n^s}}{{{m_2} - 2 - {\lambda _n}}}
\end{equation}
Hence the probability amplitude of $v_l$ which is a vertex of MC is
\begin{equation}\label{eq20}
{\alpha _{l,j}}\left( t \right) = \sum\limits_{n = 1}^N {\frac{{{a_n}\left( {{e^{i\left( {{m_1} - 2} \right)t}} - {{\rm{e}}^{{\rm{i}}{\lambda _n}t}}} \right)}}{{{m_1} - 2 - {\lambda _n}}}} ,
\end{equation}
and when $v_k$ is not  a member of the maximum clique, the probability amplitude is
\begin{equation}\label{eq21}
{\alpha _{l,j}} = \sum\limits_{n = 1}^N {\frac{{{a_n}\left( {{e^{i\left( {{m_2} - 2} \right)t}} - {{\rm{e}}^{{\rm{i}}{\lambda _n}t}}} \right)}}{{{m_2} - 2 - {\lambda _n}}}}.
\end{equation}
As $m_j - 2$ is not an eigenvalue for $j=1,2$, compare Eq.\ref{Re} with Eq.\ref{eq21} to obtain:
\begin{equation}
\sum\limits_{n = 1}^N {\frac{{{a_n}{e^{i\left( {{m_j} - 2} \right)t}}}}{{{m_j} - 2 - {\lambda _n}}}}  = 0.
\end{equation}
Let $p_{l,n}$ denote the coefficient of $\alpha_{l,j}$ at eigenvalue $\lambda_n$. Then we have the follow theorem. 
\begin{thm}
	For the first kind of ideal center graph $G_j$, $v_j,{v_l},v_k \in V ({G_j})$, $v_j$ is the center vertex, $v_l$ is a member of the maximum clique of $G_j$, and $v_k$ is not a member of the maximum clique. Then
	\[p_{l,1}>p_{k,1},\]i.e.,\[\left|\frac{1}{m_1-2-\lambda_1}\right|>\left|\frac{1}{m_2-2-\lambda_1}\right|\].
	\label{thm1}
\end{thm}
\textbf{Theorem.}\ref{thm1} is obvious since $\lambda_1>m_1-2>m_2-2$. Therefore, it is easy to determine the vertex that belongs to the maximum clique from the first kind of ideal center graph by using CTQW. 

\begin{figure}[!htbp]
	\centering\includegraphics[width=0.5\textwidth]{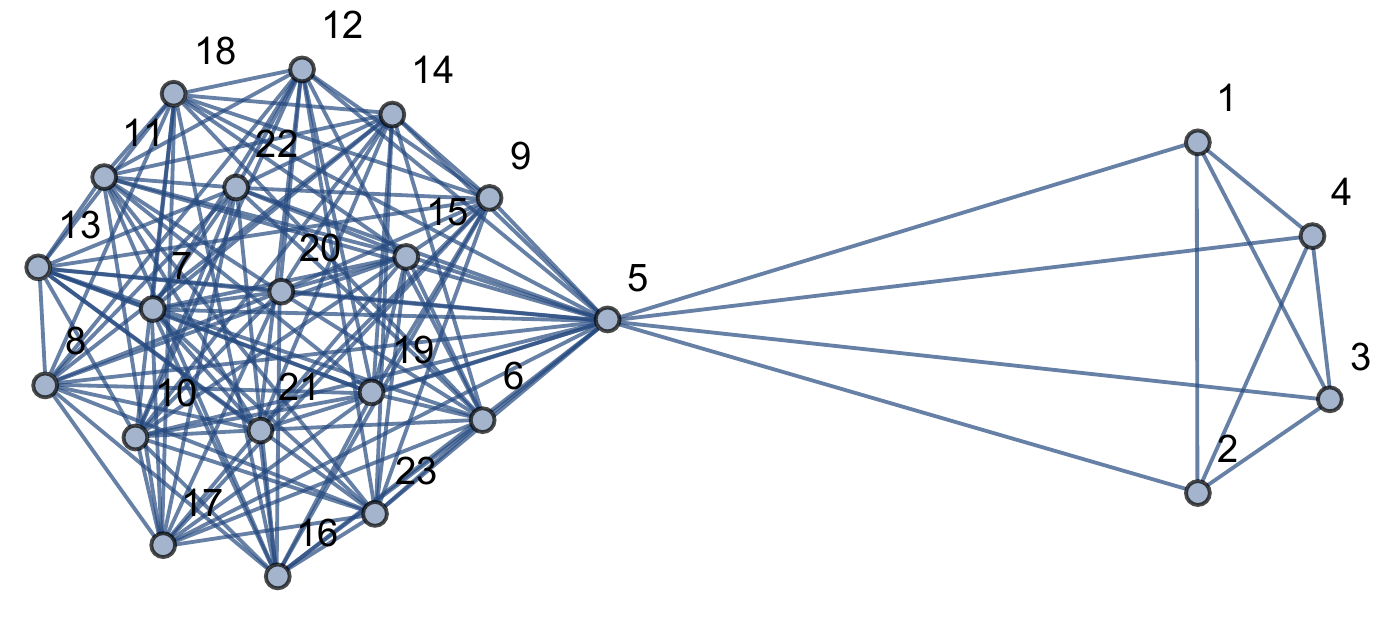}
	\caption{The second kind of ideal center graph. There has no edge that connect the vertices in the MC and the vertices not in the MC. The subgraph induced by vertices \{6,7,...,23\} is a complete multi-partite graph}
	\label{fig:6}
\end{figure}

The second kind of ideal center graph is a graph derived from the first kind of ideal center graph by replacing the second large clique with a complete multi-partite graph and keeping the maximum clique unchanged. An example graph of this type is shown in Fig.\ref{fig:6}.

The maximum clique of the graph shown in Fig.\ref{fig:6} is five. Since the subgraph induced by vertices \{6,7,...,23\} is a complete multi-partite graph, adding edges will generate cliques with sizes not less than five. 

For the second kind of ideal center graph, the number of walks can be determined by the following recursion equations:
\begin{equation}\label{eq27}
\left\{ {\begin{array}{*{20}{c}}
	{{W_{s + 1}} = \left( {{m_1} - 1} \right){F_s} + z\left( {{m_2} - 1} \right){H_s}}\\
	{{F_{s + 1}} = {W_s} + \left( {{m_1} - 2} \right){F_s}}\\
	{{H_{s + 1}} = {W_s} + z\left( {{m_2} - 2} \right){H_s}}
	\end{array}} \right.
\end{equation}
where $z$ is the number of vertices in each independent set. Comparing Eqs.(\ ref{eq27}) to Eqs.(\ref{eq13}), one can find that the solution formats of Eqs.(\ref{eq27}) are similar to the solutions of Eqs.(\ref{eq13}), just replacing $\left({m_2}-2\right)$ by $ z\left( {{m_2} - 2}\right)$ in \ref{eq21}. Therefore, for a vertex $v_l$, which belongs to the maximum clique in the graph shown in Fig.\ref{fig:6}, it is provided that:
\begin{equation}\label{eqa20}
{\alpha _{l,j}}\left( t \right) = \sum\limits_{n = 1}^N {\frac{{{a_n}\left( {{e^{i\left( {{m_1} - 2} \right)t}} - {{\rm{e}}^{{\rm{i}}{\lambda _n}t}}} \right)}}{{{m_1} - 2 - {\lambda _n}}}} ,
\end{equation}
and for a vertex which does not  belong to the maximum clique, it provides:
\begin{equation}\label{eqb20}
{\alpha _{l,j}}\left( t \right) = \sum\limits_{n = 1}^N {\frac{{{a_n}\left( {{e^{i\left( {z\left( {{m_2} - 2}\right)} \right)t}} - {{\rm{e}}^{{\rm{i}}{\lambda _n}t}}} \right)}}{{z\left( {{m_2} - 2}\right) - {\lambda _n}}}} ,
\end{equation}
Similar to \textbf{Theorem.}\ref{thm1}, the following theorem can be provided:
\begin{thm}
	For the second kind of ideal center graph $G_j$, $v_j,{v_l},v_k \in V ({G_j})$, $v_j$ is the center vertex, $v_l$ is a member of the maximum clique of $G_j$, and $v_k$ is not a member of the maximum clique. Then
	\[p_{l,1}<p_{k,1},\]i.e.,\[\left|\frac{1}{m_1-2-\lambda_1}\right|<\left|\frac{1}{z\left( {{m_2} - 2}\right)-\lambda_1}\right|,\] if 
	\[{m_1} - 2 < z\left( {{m_2} - 2} \right).\]
	\label{thm2}
\end{thm}

In the previous instances, there was no edge connecting vertices in different cliques. For general cases, multiple edges exist between vertices from distinct cliques. One such instance is exhibited in Fig.\ref{fig:8}:
\begin{figure}[!htbp]
	\centering\includegraphics[width=0.5\textwidth]{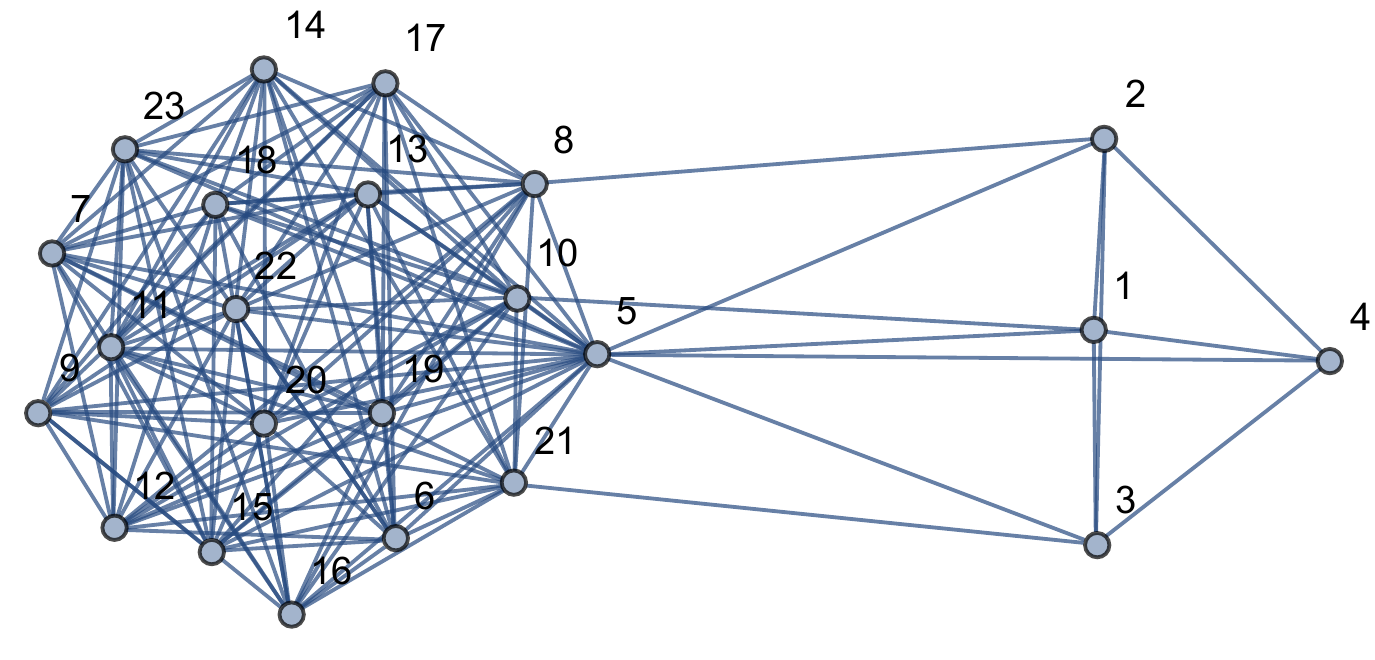}
	\caption{An exaple graph. In this configuration, the maximun clique and the non-maximum clique connect by edges (1,10),(2,8),(3,21)}
	\label{fig:8}
\end{figure}

\begin{figure}[!htbp]
	\centering\includegraphics[width=0.5\textwidth]{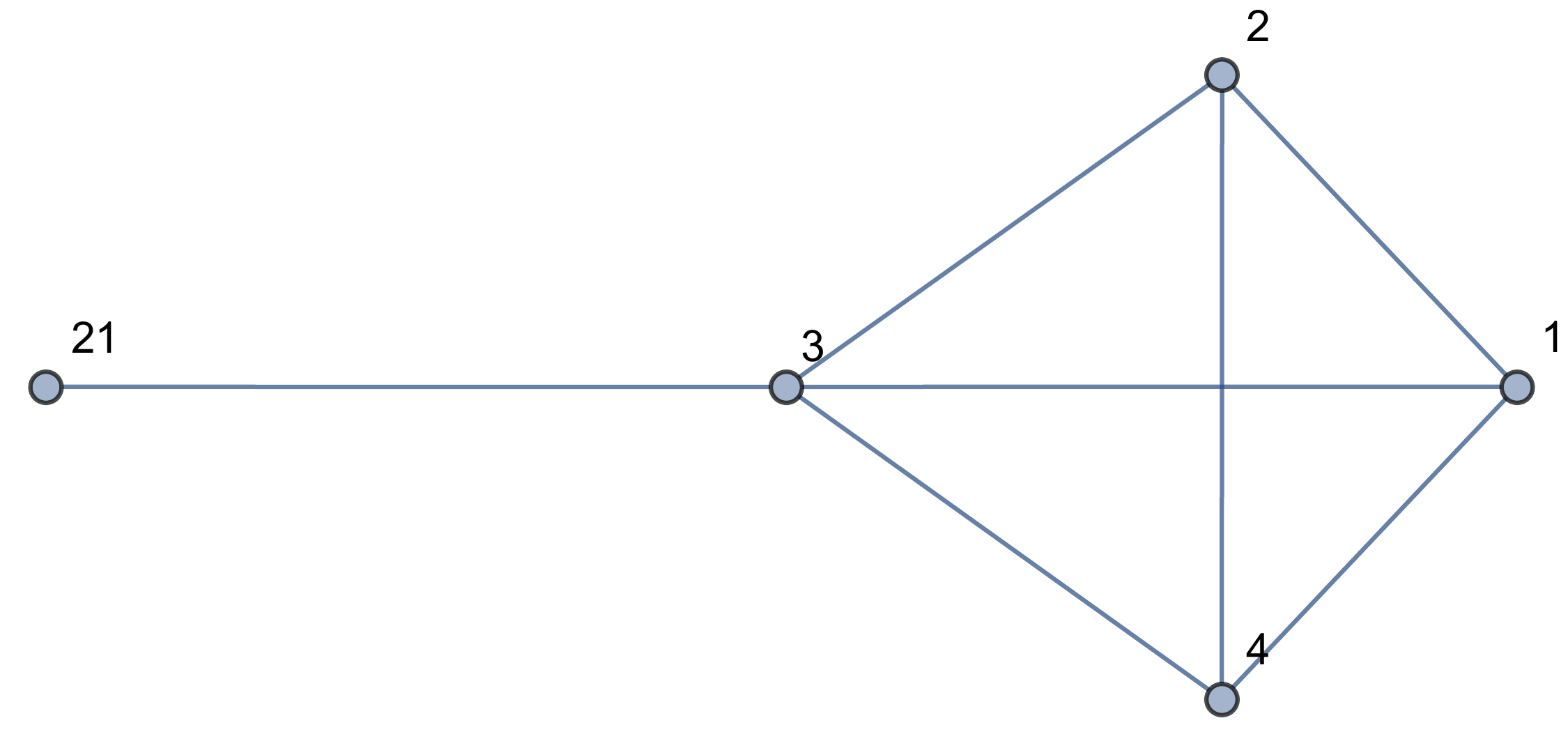}
	\caption{The center subgraph of vertex 3. The graph is induced by vertex 3 and it$\textquoteright$s neighbors except center vertex, namely \{1,2,3,4,21\}}
	\label{fig:9}
\end{figure}

For general graphs, the method of counting the number of walks to determine the CTQW is as difficult as the eigen-decomposition of the adjacency matrix. For general graphs, even when the eigen-decomposition is obtained, to directly deduce that whether a vertex is a member of the maximum clique is still unknown. An intuitive idea is to generate a center subgraph of the original graph and find the maximum clique in that center subgraph. For instance, the maximum clique attached to vertex 3 of the graph shown in Fig.\ref{fig:8} can be easily found. The center graph of vertex 3 is exhibited in Fig.\ref{fig:9}. Note that the original center vertex 5 is not included in $G_3$.

The resultant graph in Fig.\ref{fig:9} is the first kind of ideal center graph, and its maximum clique can be determined by \textbf{Theorem.}\ref{thm1}. This shows that performing CTQW on center graphs in an orderly way can help to understand the maximum clique. In this procedure, we repeatedly chose vertices and constructed associated center subgraphs. In the following section, $C(v_j)$ is used to denote the center graph of $v_j$ or the procedure of constructing the center graph of $v_j$. The procedure of deleting vertex $v_j$ in graph $G$ is frequently used and is denoted as $D(v_j,G)$. The method regarding successive selection of a vertex to construct a center graph is presented in the next section. 

\section{A recursive algorithm for finding the maximum clique by CTQW}
\begin{figure}[!htbp]
	\centering\includegraphics[width=0.5\textwidth]{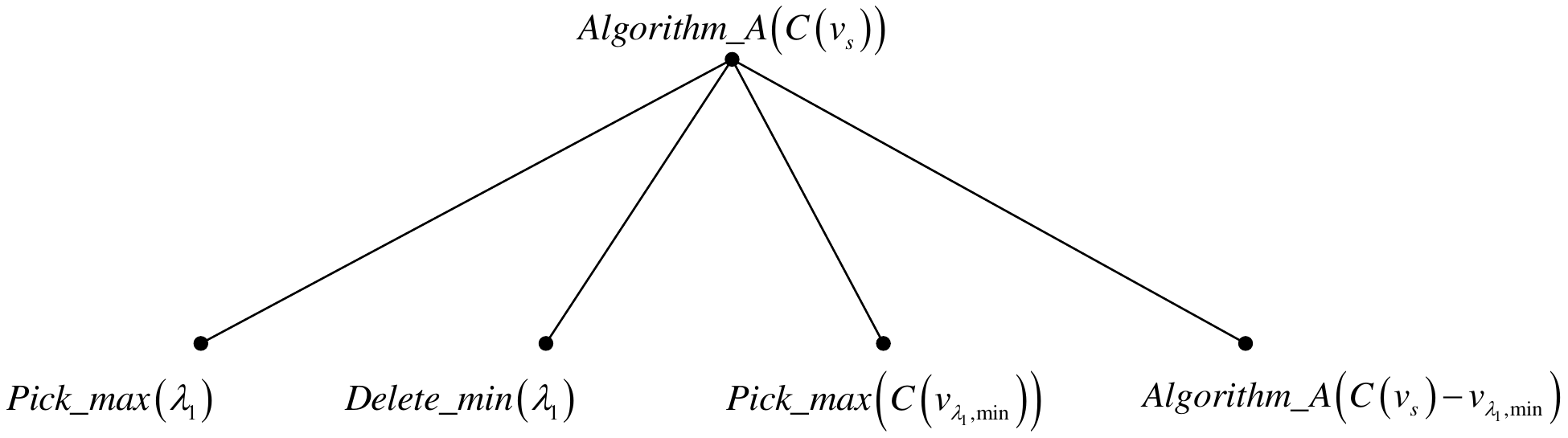}
	\caption{The recursive algorithm for finding the maximum clique. There are four sub-modules in algorithm A, and the fourth sub-module recursively calls the algorithm A on a smaller subgraph.}
	\label{fig:11}
\end{figure}

An algorithm, named algorithm A, based on eigenvectors of the adjacency matrix of the graph for finding the maximum clique is proposed in this section. The intensities of the real part (or imaginary part) of the probability amplitude are used as the critical feature in selecting probable vertices. The procedure of algorithm A is illustrated by a tree-like diagram in Fig.\ref{fig:11}. 

Algorithm A is based on the \textbf{Theorem.}\ref{thm1}, i.e., the frequency component of the largest eigenvalue $\lambda_1$ is used as the feature in choosing probable vertices belonging to the maximum clique. If the intensity $p_{k,1}$ of vertex $v_k$ is the largest among all other vertices, then it would be chosen as a probable member of the maximum clique. The root of the tree-like diagram denotes the algorithm, and the leaf node denotes the sub-modules of the algorithm.

As shown in Fig.\ref{fig:11}, there are four cases (sub-modules) in algorithm A, and the last is the recursive process. The first one is the sub-algorithm named $Pick\_max$ in which the strategy of choosing the vertex with the largest intensity at frequency $\lambda_1$ in amplitude of CTQW is employed. The second one is the sub-algorithm named $Delete\_min$ in which the strategy of deleting the vertex with the weakest intensity at frequency $\lambda_1$ in amplitude of CTQW is employed. Then, we wanted to delete the vertex with the weakest intensity at frequency $\lambda_1$, and this vertex is denoted as $v_{\lambda_1,min}$. Since that vertex may be a member of the maximum clique, the third sub-module $Pick\_max \left( {C\left( {{v_{\lambda_1,min}}} \right)} \right)$ is applied to find the maximum clique of $v_{\lambda_1,min}$. The last sub-module is to recursively call the algorithm A. Note that the size of the graph decreases by one since a vertex is deleted in the fourth module. The maximum clique of $C(v_s)$ is the largest clique found in these four cases. The sub-modules $Pick\_max$ and $Delete\_min$ are presented in the following charts.

\begin{table}[H]  
	\caption{ The step of Pick\_max}  
	\label{alg:Framwork}  
	\begin{algorithmic}[1]  
		\Require  
		Center graph $G_s$, center vertex $v_s$
		\Ensure  
		Clique C;  
		\State  Delete vertex $v_s$ in $G_s$ and for every other vertex $v_l$, let $C_l=\{v_s,v_l\}$, find the central graph of $v_l$, and denote as $G_l$, $v_s$ $\longleftarrow$ $v_l$;
		\label{code:1}  
		\State 	Do eigen-decomposition on adjacent matrix of $G_l$, The largest eigenvalue is $\lambda_1$, corresponding eigenvector is $\bf{x_1}$ and $x_1^l$ denotes the $l$-th component of $\bf{x_1}$;  
		\label{code:2}  
		\State Calculate the intensity of amplitude of every vertex $v_k$ at the maximum frequency $\lambda_1$, denoted as $p_k$, $p_k=x_1^l x_1^k$; 
		\label{code:3}  
		\State Add vertex $v_{1,max}$ which has the largest intensity at frequency $\lambda_1$ to $C_l$; Delete $v_l$ and construct center graph $C(v_{1,max})$, if size of $C(v_{1,max})$ is not 1, then $v_l$$\longleftarrow$ $v_{1,max}$ and turn to step.\ref{code:2}, else turn to step.\ref{code:5}
		\label{code:4}  
		\State Select the clique of max size in $C_l$ as $C$.
		\label{code:5}
	\end{algorithmic}  
\end{table}  
\begin{table}[H]  
	\caption{ The step of Delete\_min}  
	\label{alg:Step}  
	\begin{algorithmic}[1]  
		\Require  
		Center graph $G_s$, center vertex $v_s$
		\Ensure  
		Clique C;  
		\State If $G_s$ is a complete graph, then return all vertexes of $G_s$ as clique $C$. 
		\label{code:6}  
		\State 	Do eigen-decomposition on adjacent matrix of $G_s$, The largest eigenvalue is $\lambda_1$, corresponding eigenvector is $\bf{x_1}$ and $x_1^l$ denotes the $l$-th component of $\bf{x_1}$;
		\label{code:7}  
		\State Delete vertex $v_{1,min}$ which has the smallest intensity at frequency $\lambda_1$, turn to step.\ref{code:6}.
		\label{code:8}   
	\end{algorithmic}  
\end{table}  

Algorithm A has polynomial complexity. To observe this, we must solve the recursion of algorithm A. Let $T(n)$ denote the complexity of algorithm A, and $Y_1(n)$, $Y_2(n)$, and $Y_3(n)$ denote the complexity of the corresponding submodules. Then the following recursion is satisfied.
\begin{equation}\label{recursion1}
T\left( n \right) = {Y_1}\left( n \right) + {Y_2}\left( n \right) + {Y_3}\left( n \right) + T\left( {n - 1} \right)
\end{equation}
As the complexities of $Y_1(n)$, $Y_2(n)$, and $Y_3(n)$ are all $O(n^4)$, Eq.\ref{recursion1} can be reduced to: 
\begin{equation}\label{recursion2}
T\left( n \right) = O\left( {{n^4}} \right) + T\left( {n - 1} \right)
\end{equation}
Then from Eq.\ref{recursion2}, it provides that 
\begin{equation}\label{complexity}
T\left( n \right) = O\left( {{n^5}} \right)
\end{equation}

We have experimented on random graphs with different edge connecting probabilities varying from 0.3 to 1 on classical computers. In our experiments, counter examples, where algorithm finds a sub-maximal clique, has not yet been found. However, counter examples can be elaborately constructed. We will illustrate an approach for designing such a graph in the next section.

\section{Probable counter examples of algorithm A } 
\begin{figure}[!htbp]
	\centering\includegraphics[width=0.5\textwidth]{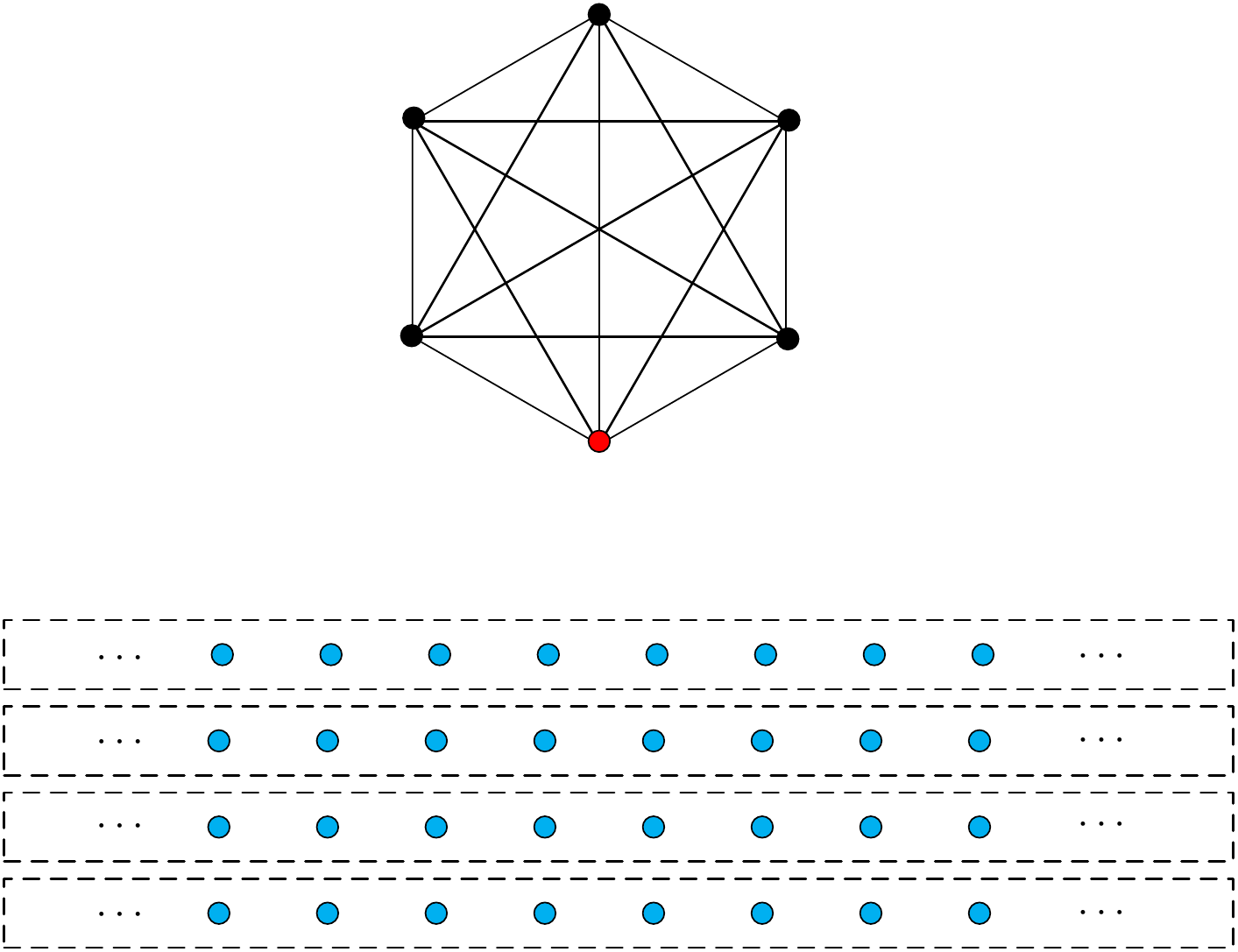}
	\caption{Base graph. The graph contains a complete subgraph and a complete multi-partite subgraph. The number of independence set is not larger than the clique number, the number of vertices in the complete multi-partite subgraph is as great as possible.}
	\label{fig:20}
\end{figure}

We will present an approach for constructing probable counter examples of algorithm A in this section. Let $W_s^{v_j}$ denote the number of walks of length $s$ from the center vertex to the vertex ${v_j}$. From the previous section,
\begin{equation}
W_s^{{v_j}} = \sum\limits_{k = 1}^N {a_k^{{v_j}}\lambda _k^s}.
\end{equation}
Where, $a_k^{v_j} = \left\langle s \right|\left. {{\lambda _k}} \right\rangle \left\langle {{\lambda _k}} \right|\left. {{v_j}} \right\rangle $. Therefore, the amplitude of CTQW is 
\[{p_{{v_j}}} = \sum\limits_{k = 1}^N {a_k^{{v_j}}{e^{i{\lambda _k}t}}}.\]
In algorithm A, $a_1^{v_j}$ of different vertices at the largest eigenvalue $\lambda_1$ are compared. For a large enough $s$, $a_1^{v_j}$ and $a_1^{v_h}$, $a_1^{v_j}>a_1^{v_h}$ if and only if $W_s^{v_j}>W_s^{v_h}$. This implies that if algorithm A is invalid for some graph $G$, then every member of the maximum clique has a neighbor $v_h$ that has the largest number of walks $W_s^{v_h}$. In this case, no matter which vertex belonging to the maximum clique is chosen, $v_h$ will be chosen in some layer of the recursion and algorithm A spontaneously fails. For simplicity, assume all such subgraphs that are adjacent to every member of the maximum clique are the same, and it is a complete multi-partite graph with a degree far larger than the size of the MC. For clarity, we propose a different kind of graph named as a base graph. A schematic diagram of a base graph is shown in Fig.\ref{fig:20}.

The base graph consists of two elementary sub-graphs, one is the maximum clique at the upper of Fig.\ref{fig:20}, the other is a complete multi-partite graph containing all the light blue vertices in Fig.\ref{fig:20}. The vertex set in the same dashed box is not adjacent pairwise, i.e., it is an independent set. Also, vertices from different dashed boxes are fully adjacent. Edges can be added between the vertices of the maximum clique and the complete multi-partite graph. First, a vertex, the red vertex in this instance, is chosen as the center vertex. Secondly, every combination of three vertices except the center vertex in the maximum clique are connected to $q \cdot z\left( {qz >\omega \left( G \right)-3,q + 4 < \omega \left( G \right)} \right)$ common vertices from at least two partites  of the complete multi-partite graph, where $q$ is the number of partites, $z$ is the number of vertices in each independent set and $\omega \left( G \right)$ is the clique number. After the two procedures adding edges, every three vertices of the maximum clique have a common complete multipartite graph as their neighbor. The condition ${qz >\omega \left( G \right)-3}$ ensures that a non-maximum clique vertex will be chosen in the procedure of algorithm A, namely algorithm A failed in this situation . 

Although algorithm A is not universal, it divides all graphs into two classes. The maximum clique of the first class of graphs can be accurately determined by algorithm A but the second class of graphs cannot. Therefore, if an algorithm exists that can crack the second class of graphs, then the problem can be overcome. Algorithm B is designed to improve the performance of algorithm A for the second class of graphs, and algorithm C is derived from algorithm B by removing recursions. We will further describe them in the next section.

\section{Variational frequency selection algorithm for finding clique}
\begin{figure}[!htbp]
	\centering\includegraphics[width=0.5\textwidth]{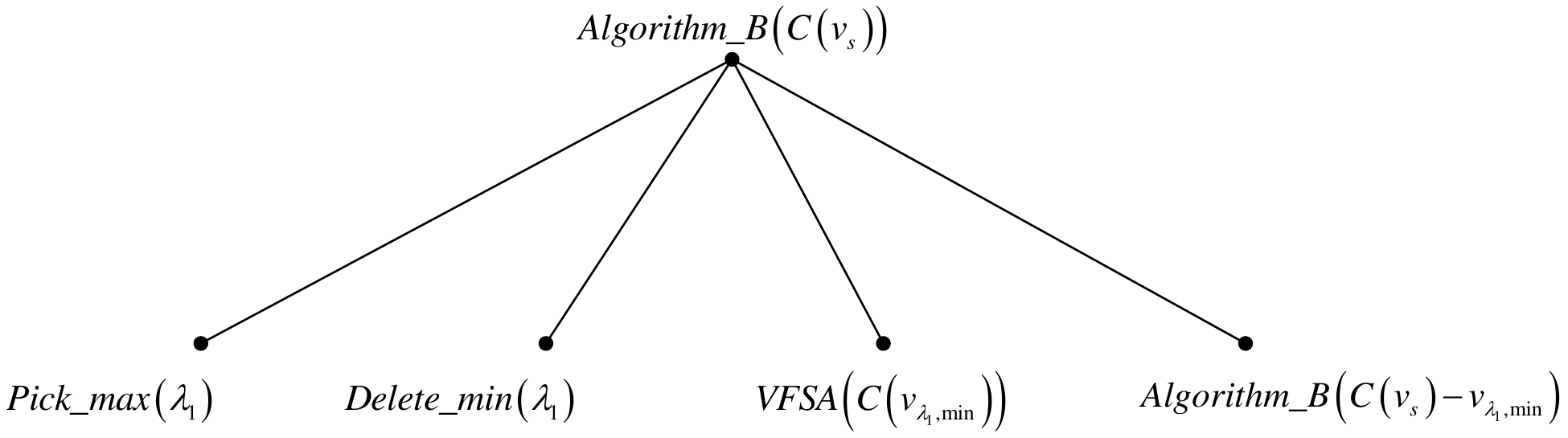}
	\caption{The procedure of algorithm B. The algorithm has four sub-modules and only the third sub-module is different with the algorithm A.}
	\label{fig:13}
\end{figure}

In algorithm B, we give greater attention to contextual information. We assume that $v_s$ is a vertex with the smallest intensity at the largest frequency, and we want to find a maximum clique attached to $v_s$. In the center graph of the original center vertex, $v_s$ has the smallest intensity at the largest frequency; however, $v_s$ has a considerable intensity at some frequency $\lambda_j$, then the other members of the maximum clique are more likely to occur at frequency $\lambda_j$. Then a vertex $v_{ref}$ with the largest intensity at frequency $\lambda_j$ except $v_s$ is chosen, and $v_{ref}$ is called the reference vertex. The CTQW on the center graph of $v_s$ and the location of vertex $v_{ref}$ is used to determine the next reference vertex. The $v_{ref}$ will become the new center vertex and a new reference vertex is used to find the next reference vertex. The counter examples of algorithm B is not found or not elaborately constructed. To prove algorithm B is valid for the second class of graphs is beyond this work. The procedure of algorithm B is illustrated in Fig.\ref{fig:13}.

The third module is named VFSA (variational frequency selection algorithm) for finding clique. The input of VFSA are $C(v_{\lambda_1,min})$ and $v_{ref}$. Where $C\left( v_{\lambda_1,min} \right)$ is the center graph of vertex $v_{\lambda_1,min}$ which signifies the vertex with the weakest intensity at the frequency ${\lambda _1}$. Let $f_{ref}$ denote the frequency where $v_{\lambda_1,min}$ takes the largest intensity in graph $C(v_s)$. And where vertex $v_{ref}$ is the auxiliary adjacent vertex of $v_{\lambda_1,min}$ and has the largest intensity except $v_{\lambda_1,min}$ in frequency $f_{ref}$. Vertex $v_{ref}$ acts as the central vertex in the subsequent procedure. Since $f_{ref}$ and $v_{ref}$ can be determined when $v_{\lambda_1,min}$ is given, we regard them as two implicit parameters and do not show them on the tree-like diagram \ref{fig:13}. The the steps of VFSA are listed in the following table.

\begin{table}[H]  
	\caption{ The step of VFSA}  
	\label{alg:Procedure}  
	\begin{algorithmic}[1]  
		\Require  
		Center graph $G$, center vertex $v_s$, next center vertex $v_{ref}$
		\Ensure  
		Clique $C$;  
		\State If $G$ is a complete graph, then return all vertexes of $G$ as clique $C$, if not, add $v_s$ to clique $C$, turn to step.\ref{code:10}
		\label{code:9}  
		\State 	Do eigen-decomposition on adjacent matrix of $G$, The largest eigenvalue is $\lambda_1$, corresponding eigenvector is $\bf{x_1}$ and $x_1^l$ denotes the $l$-th component of $\bf{x_1}$. Find a frequency $f_{ref}$ such that vertex $v_{ref}$ can take the largest intensity among all frequencies.
		\label{code:10}  
		\State 	Find vertex $v_{ref}^\prime$ which has the largest intensity at frequency $f_{ref}$ in the neighbors of vertex $v_{ref}$
		\label{code:11}  
		\State Delete $v_s$ and let $v_s$$\longleftarrow$$v_{ref}$, $v_{ref} \longleftarrow v_{ref}^\prime$, $G_s \longleftarrow C(v_{ref})$, turn to step.\ref{code:9}
		\label{code:12} 
	\end{algorithmic}  
\end{table}  

In both algorithms A and B, the recursion is utilized to eliminate probable interference from low connectivity vertices. However, since a vertex will be removed in the recursion, not all of the maximum cliques will be found when there are multiple maximum cliques attached to the same vertex. For this case, we transform algorithm B to an algorithm without recursion. The recursion-less algorithm, algorithm C, calls VFSA for every vertex of the given graph $N-1$ times. The time complexity of algorithm C is $O(N^5)$, as the complexity of VFSA is $O(N^4)$ time. 
\begin{figure}[!htbp]
	\centering\includegraphics[width=0.5\textwidth]{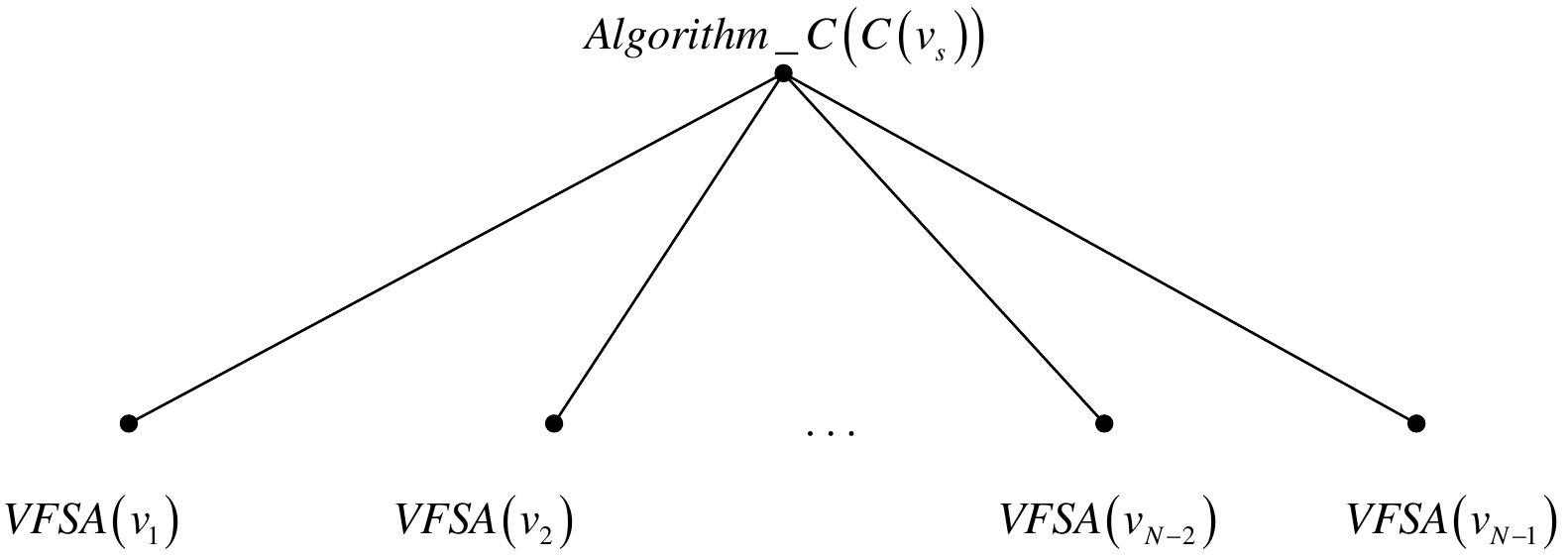}
	\caption{Algorithm C.}
	\label{fig:113}
\end{figure}

\section{Conclusion}
In this theme, we show that the clique structure of a graph is related to the CTQW. For some ideal graphs, the frequency intensity of the probability amplitude of CTQW is a good feature to directly speculate whether a vertex is a member of the maximum clique or not. As the frequencies of CTQW are the eigenvalues of the adjacency matrix, the clique structure is related to the eigenvalues and the corresponding eigenvectors. For general graphs, this feature is not so obvious, and one cannot directly find the maximum clique. To reveal the hidden maximum clique, we propose two recursive algorithms, algorithm A and algorithm B, using CTQW with $O(N^5)$ time complexity to find the maximum clique on graphs. We further transform algorithm B to algorithm C for graphs with multiple maximum cliques. It seems that the algorithm A is valid for random graphs via numerical experiments, but counter examples can be elaborately constructed. For such graphs whose maximum clique cannot be found by algorithm A, we propose algorithm B. To understand whether algorithm B can be used for all graphs is beyond this article but will be undertaken in future work.

\section*{Acknowledgments}
This work is supported by the National Natural Science Foundation of China (Grant No. 61170321,61502101,61871120,61802002), Natural Science Foundation of Jiangsu Province, China (Grant No. BK20140651), Natural Science Foundation of Anhui Province, China (Grant No. 1608085MF129), Research Fund for the Doctoral Program of Higher Education (Grant
No. 20110092110024), Foundation for Natural Science Major Program of Education Bureau of Anhui Province (Grant No. KJ2015ZD09) and the open fund of Key Laboratory of Computer Network and Information Integration in Southeast University, Ministry of Education, China (Grant No. K93-9-2015-10C).

\bibliographystyle{unsrtnat}
\bibliography{ref}

\end{document}